\documentclass{aa}
\pdfoutput=1

\usepackage{twoopt}
\bibpunct{(}{)}{;}{a}{}{,}

\makeatother

\usepackage{graphicx}
\usepackage{footnote}
\usepackage{amsmath}
\usepackage{amssymb}
\usepackage{pdflscape}
\usepackage{url}
\usepackage[varg]{txfonts}
\usepackage{bm}
\usepackage{booktabs}
\usepackage{svg}
\usepackage{hyperref}
\hypersetup{
    colorlinks=true,
    citecolor=blue,
    linkcolor=blue,
    urlcolor=blue
}
\usepackage{orcidlink}

\let\orcid\relax
\newcommand{\orcid}[1]{\href{https://orcid.org/#1}{\includegraphics[width=0.8em,height=0.8em]{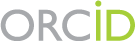}}}

\DeclareGraphicsExtensions{.pdf,.png,.jpg}

\begin{document}

   \title{$J$ and $H$ band sky brightness measurements from polar day to polar night at Dome A, Antarctica}

    \author{
        Jinji Li
        \orcidlink{0009-0000-6955-0594}
        \inst{1}
        \and
        Bin Ma
        \orcidlink{0000-0002-6077-6287}
        \inst{1,2\thanks{Corresponding author: mabin3@mail.sysu.edu.cn}}
        \and
        Haonan Yang
        \orcidlink{0009-0007-2408-9221}
        \inst{1}
        \and
        Pu Lin
        \orcidlink{0009-0007-2850-9908}
        \inst{1}
        \and
        Zhongnan Dong
        \orcidlink{0009-0003-5592-3734}
        \inst{1,3}
        \and
        Michael C.~B. Ashley
        \orcidlink{0000-0003-1412-2028}
        \inst{4}
        \and
        Lu Feng
        \orcidlink{0000-0002-1175-3935}
        \inst{3}
        \and
        Yi Hu
        \orcidlink{0000-0003-3317-4771}
        \inst{3}
        \and
        Zhaohui Shang
        \orcidlink{0000-0002-6796-124X}
        \inst{3}
        \and
        Yun Shi
        \orcidlink{0009-0003-0141-5793}
        \inst{1}
        \and        
        Shijie Sun
        \orcidlink{0000-0003-2775-3523}
        \inst{3}
        \and
        Xu Yang
        \orcidlink{0000-0003-4147-8759}
        \inst{3}
    }
    
    \institute{
        School of Physics and Astronomy, Sun Yat-sen University, Zhuhai 519082, China
        \and
        CSST Science Center for the Guangdong-Hong Kong-Macau Greater Bay Area, Zhuhai 519082, China
        \and
        National Astronomical Observatories, Chinese Academy of Sciences, Beijing 100101, China
        \and
        School of Physics, University of New South Wales, Sydney, NSW 2052, Australia
    }

   \date{Received XXX; accepted XXX}

    \abstract
    {The near-infrared (NIR) sky brightness is a fundamental parameter for evaluating the performance of ground-based infrared observatories. 
    }
    {Dome~A on the Antarctic plateau offers exceptional atmospheric conditions, yet its NIR sky background has not been continuously monitored. 
    We present the first continuous $J/H$-band measurements of the sky background at Dome~A from polar day to polar night, and characterize their median levels and temporal variability.
    }
    {The Antarctic Infrared Binocular Telescope (AIRBT), operating in the $J$ and $H$ bands, obtained continuous fixed-pointing observations from February to May 2024, which were used to measure the NIR sky background.}
    {The median sky brightness is $5.2/2.9$ and $15.3/13.4~\mathrm{mag~arcsec^{-2}}$ in $J/H$ bands during daytime and nighttime, respectively. 
    The twilight--nighttime boundaries occur at solar elevations of $-9.3^\circ$ in $J$ and $-7.4^\circ$ in $H$. 
    At the same solar elevation, the NIR sky background during the polar night is darker by about $0.1$ and $0.4~\mathrm{mag~arcsec^{-2}}$ in the $J$ and $H$ bands compared with the period of regular day--night alternation.
    During the polar-night period, the nighttime sky brightness in the $H$ band shows 
    a more evident association with the sunspot number, while the corresponding trend in 
    the $J$ band is weaker.}
    {These results reveal systematic differences in sky background between polar and non-polar environments and between polar night and regular day--night cycles. The measured sky brightness may be elevated, as the observations were conducted near solar maximum, highlighting the importance of long-term monitoring across the solar cycle.}


    \titlerunning{$J/H$-band sky brightness at Dome~A}

   \maketitle
   \nolinenumbers

\section{Introduction}

The sky brightness is a fundamental parameter for evaluating the quality of an astronomical site, as it limits the sensitivity and photometric precision of observations.
Across the entire electromagnetic spectrum, from the far-ultraviolet to the far-infrared, the physical origins of the sky background have been extensively investigated \citep{leinert19981997}.
At different wavelengths, the dominant atmospheric and radiative processes vary, leading to significant differences in the sky background levels among observing sites \citep{sivanandam2012characterizing}.

In the near-infrared (NIR), the sky background is typically several magnitudes brighter than in the optical \citep{birch2022ingaas}.
At Mauna Kea, for example, the typical $J$-band sky brightness is $\sim15.6~\mathrm{mag~arcsec^{-2}}$, compared with a $V$-band sky brightness of $\sim21.9~\mathrm{mag~arcsec^{-2}}$ \citep{patat2004night, sanchez2008night}.
This large difference arises because the NIR sky background is dominated by strong airglow emission from hydroxyl (OH) molecules \citep{maihara1993observations}.
Excited OH radicals reside in the mesosphere, at altitudes of 80--105~km, and are primarily generated through the chemical reaction:
\begin{equation}
\begin{aligned}
   \mathrm{H + O_3 \rightarrow OH^* + O_2},
\end{aligned}
\label{oh_emission}
\end{equation}
The excited OH radicals subsequently undergo radiative cascades to lower energy states, producing the characteristic airglow emission features observed in the NIR bands \citep{bates1950photochemistry}.

The NIR sky background exhibits substantial temporal variability.
On short timescales (seconds to minutes), hydroxyl airglow shows fluctuations in both intensity and spatial distribution, driven largely by variations in OH abundance induced by atmospheric gravity waves \citep{ramsay1992non, swenson1994oh}. 
On longer timescales (annual to decadal), mesospheric OH airglow exhibits a well-established positive correlation with the solar cycle, with its intensity increasing toward solar maximum \citep{pertsev2008response}.
The properties of the NIR sky background also vary significantly from site to site, particularly in their seasonal behavior. 
At Calar Alto and Mount Abu, the $J$- and $H$-band sky backgrounds remain relatively stable throughout the year, whereas the $K$ band exhibits pronounced seasonal variability \citep{sanchez2008night, prajapati2023near}.  
At Calar Alto, for instance, the difference between the darker winter sky and the brighter summer sky in the $K$-band can reach $\sim0.8~\mathrm{mag~arcsec^{-2}}$.
In contrast, Mauna Kea shows noticeable seasonality primarily in the $J$ band, with a gradual darkening trend from late spring through winter, particularly during the second half of the night \citep{roth2016measurements}.
Despite these site-to-site differences, a common characteristic of the $J$-band sky is that it typically requires $\sim$20--30\,\% of the night to reach a relatively stable level, where nighttime is defined as a solar elevation below $-12^\circ$ \citep{vaduvescu2004strategies, sanchez2008night, roth2016measurements, tatarnikov2024brightness}.

Over the past three decades, site-testing efforts have established the Antarctic plateau as the darkest ground-based site for infrared astronomy \citep{burton2010astronomy}. 
Early measurements at the South Pole revealed exceptionally low NIR backgrounds.
For example, \citet{ashley1996south} and \citet{nguyen1996south} reported that the 2.3--2.5~$\mu$m $K_{\rm dark}$ window exhibits sky fluxes of only $\sim$50--250~$\mu\mathrm{Jy~arcsec^{-2}}$ (17.7--16.0~$\mathrm{mag~arcsec^{-2}}$), up to two orders of magnitude lower than those measured at temperate sites. Subsequent observations confirmed similarly low and stable backgrounds over the 1--5~$\mu$m range during the Antarctic winter \citep{phillips1999near}.
Parallel mid-infrared studies further highlighted the advantages of the Antarctic atmosphere. \citet{smith1998mid} showed that the 10 and 20~$\mu$m backgrounds at the South Pole are markedly darker and more stable than at mid-latitude observatories.
Extending such measurements to longer infrared wavelengths, \citet{walden2005first} reported the first $M$, $N$, and $Q$-band results from Dome~C, finding exceptionally stable conditions with $N$-band sky brightness frequently below 50--60~$\mathrm{Jy~arcsec^{-2}}$.

Dome~A in Antarctica offers exceptional free-atmosphere seeing, extremely low precipitable water vapor, a high clear-sky fraction, and a continuously observable polar night \citep{shang2020astronomy,ma2020night,yang2021cloud}.
These unique atmospheric and environmental conditions make Dome~A one of the most suitable locations on Earth for optical and infrared astronomy. 
Optical observations indicate that the sky brightness in the $B$, $V$, and $R$ bands at Dome~A is comparable to that at leading mid-latitude observatories, while the $i$-band sky is darker by about 0.4--0.5~$\mathrm{mag~arcsec^{-2}}$ \citep{yang2017optical, zou2010sky}.

In the NIR, sky background measurements at Dome~A have only become available in recent years. 
\citet{zhang2023sky} used the Near-Infrared Sky Brightness Monitor (NISBM), 
a three-channel near-infrared radiometer based on InGaAs detectors, to obtain
the first $J$, $H$, and $K_s$ observations in early 2019.
Because the daytime $J/H$ measurements relied on a neutral-density filter that was later found to be improperly mounted, only the daytime $K_s$ data were considered reliable.
During April~15--17, the first three days following the onset of the polar night, their measurements yielded zenith sky brightness levels of 15.35--16.0~$\mathrm{mag~arcsec^{-2}}$ in $J$, 13.92--14.85~$\mathrm{mag~arcsec^{-2}}$ in $H$, and 14.61--16.24~$\mathrm{mag~arcsec^{-2}}$ in $K_s$, indicating that Dome~A is comparable to Mauna Kea in the $J$ and $H$ bands and substantially darker in $K_s$.
More recently, \citet{yang2025first} using a 150~mm infrared telescope carried out daytime $J$-band photometric observations in 2024, finding a zenith sky brightness of $\sim$5.2--5.8~$\mathrm{mag~arcsec^{-2}}$ for solar elevations between $27^\circ$ and $10^\circ$.
While these earlier studies provided quantitative measurements on the NIR sky background at Dome~A, they were limited either to short observing windows or to daytime-only measurements, and therefore did not characterize the continuous evolution of the $J$- and $H$-band sky background over the full annual cycle.

The Antarctic Infrared Binocular Telescope (AIRBT; \citealp{dong2025antarctic}) was installed at Dome~A in 2023 to enable long-duration infrared time-domain observations as well as sky background measurements.
In 2023, $H$-band observations were obtained; however, the data were affected by defocus, which limited data quality, and reduction is still in progress.
We therefore focus in this work on the 2024 dataset.
During its 2024 observing campaign, AIRBT obtained $J$- and $H$-band measurements of the sky background from the late polar day to the early polar night, providing the first season-long characterization of NIR sky brightness and its temporal variability at this site.

The paper is organized as follows:
Sect.~\ref{sec:instrument} introduces the AIRBT telescope, the observing strategy, and the data acquisition. 
Sect.~\ref{sec:data} describes the data reduction and sky brightness calibration procedures. 
The temporal behavior of the $J$- and $H$-band sky background is presented in Sect.~\ref{sec:results}. 
Sect.~\ref{sec:discussion} discusses the influence of solar and geomagnetic activity, the Moon and aurora. 
Finally, Sect.~\ref{sec:summary} summarizes our findings.
All the magnitudes listed in this article are in the Vega system.

\section{Observations and data}
\label{sec:instrument}

\subsection{Telescope overview}

AIRBT comprises two optical tube assemblies (OTAs) designed for NIR sky background monitoring and time-domain astronomical studies. It is located at Dome~A, Antarctica. 
The telescope has an aperture of 15~cm and a focal ratio of $f/3$, and is equipped with two InGaAs detectors, each with a $640\times512$ pixel array and a pixel size of 15~$\mu$m.
Each pixel maps to a square of side 6.84~arcsec on the sky, resulting in a field of view of $\sim1.22\times0.98~\mathrm{deg^{2}}$.
Since the InGaAs detectors employed by AIRBT have a wavelength cutoff near 1.65~$\mu$m, the effective $H$-band filter (1.48--1.65~$\mu$m) covers just over half of the 2MASS $H$-band.
This mismatch in bandpass introduces a minor systematic offset in sky background measurements. 
Following the calculations of \citet{li2024preliminary}, the AIRBT $H$-band sky brightness is expected to appear brighter by 0.11~$\mathrm{mag~arcsec^{-2}}$ relative to the 2MASS $H$ band.
We do not apply a correction for this offset, but we do note it later when comparing with results in the standard 2MASS $H$ band.

AIRBT was installed at Dome~A, Antarctica, in January 2023 by the 39th China National Antarctic Research Expedition (CHINARE~39).
Following maintenance performed by CHINARE~40 in January 2024, AIRBT began conducting simultaneous observations in the $J$ and $H$ bands.
All power supply and data transfer were provided by the PLATeau Observatory for Dome~A (PLATO-A; \citealp{lawrence2009plato, 2010EAS....40...79A}).

\subsection{Observation strategy and data acquisition}

In 2024, a mount failure restricted the telescope to a limited range of motion. 
The pointing was therefore set to $344.8^\circ$ in azimuth and $84.39^\circ$ in altitude, representing a compromise between NIR sky background monitoring and time-domain observations. 
For sky background measurements, a pointing close to the zenith is preferred because zenith brightness is the standard site-testing metric. All measurements are corrected to the zenith using a standard airmass scaling.
For time-domain studies, however, a field farther from the South Celestial Pole corresponds to a higher declination and therefore covers a larger region of the sky each day due to the Earth’s rotation.

The image acquisition was fully automated, with gain mode and exposure time adjusted in real-time according to the sky brightness.
This ensures that the images remain unsaturated under varying illumination conditions. Because the telescope pointing is fixed, the exposure settings must also balance observing efficiency against the point-spread function (PSF) elongation caused by stellar trailing, which limits the maximum exposure time to 3~s; we typically used 2~s.
With these constraints jointly optimized, the system is able to acquire continuous data throughout the entire 24-hour cycle.

In 2024, AIRBT began observations at Dome~A in January and continued operating until 1~May, when a power failure occurred in the PLATO-A system.
The initial period from January to late February was mainly used for testing and commissioning.
After this commissioning phase, the telescope entered a stable observing mode on 23~February~2024, from which continuous $J$- and $H$-band sky background measurements were obtained.
At the beginning of this observing window, the solar elevation at Dome~A ranged from $\sim19.6^\circ$ to $0.2^\circ$.
By 1~May, it had decreased to between $\sim-5.9^\circ$ and $-25.1^\circ$, corresponding to conditions from the late polar day to the early polar night. 
These observations therefore cover a broad range of solar elevations and provide a unique dataset of the NIR sky background at Dome~A over this period. 
In addition to sky background measurements, the AIRBT dataset also enables time-domain studies.
A detailed analysis of the resulting light curves will be presented in a forthcoming work (Yang, in prep.).

\section{Data reduction}
\label{sec:data}

\subsection{Preprocessing}

\begin{figure}
\resizebox{\hsize}{!}{\includegraphics{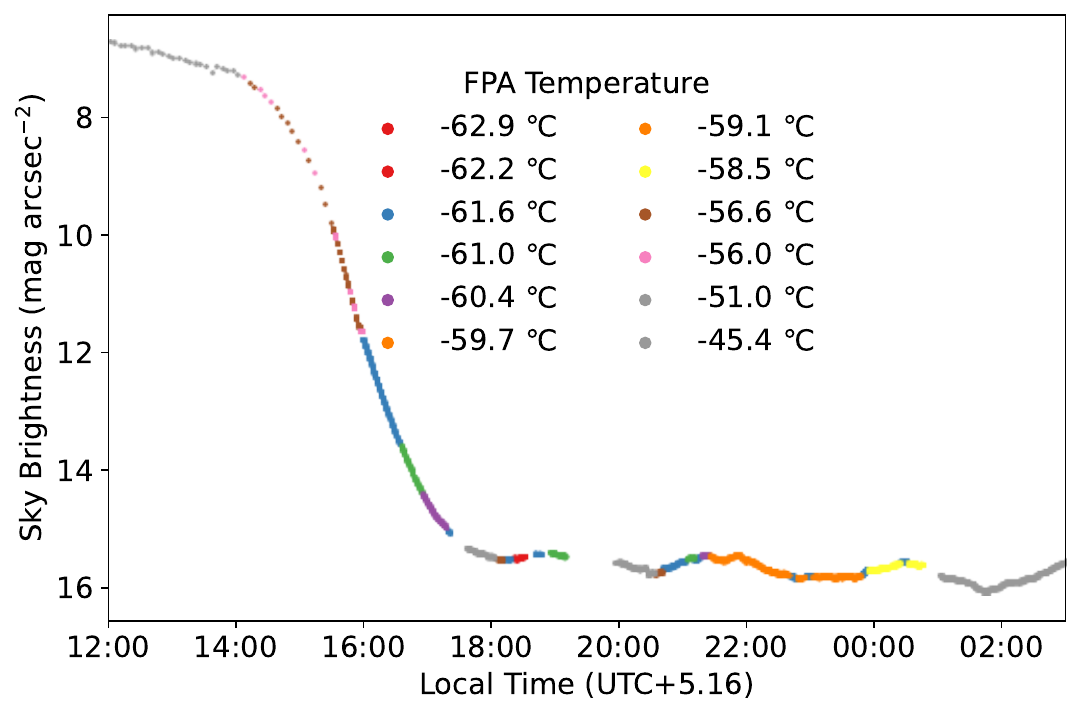}}
\caption{
Demonstration of the robustness of the warm-pixel--scaled dark-current correction.
Sky brightness during the first half of 16 April 2024 is shown as the FPA temperature varies between $-45.4^\circ$C and $-62.9^\circ$C.
Despite these large temperature changes, the derived sky brightness evolves smoothly with no discontinuities across temperature transitions.
}
\label{fig:sky_conti}
\end{figure}

A key step in the data reduction is the accurate removal of dark current.
Because AIRBT has neither a lens cap nor a dome and operates fully unattended, on-site dark frames cannot be acquired during the observing campaign.
Instead, we relied on laboratory dark-current measurements.
Since the detector is cooled by a thermoelectric cooler (TEC), variations in the ambient temperature can cause fluctuations in the detector temperature, leading to changes in the dark-current level.
For example, a focal plane array (FPA) temperature change from $-43.1^\circ$C to $-61.6^\circ$C alters the dark current by $\sim170~e^{-}\,\mathrm{s^{-1}\,pix^{-1}}$. 
This dark-current variation is negligible ($<0.004\%$) compared with the large sky-background decrease from  $\sim5.08\times10^6~e^{-}\,\mathrm{s^{-1}\,pix^{-1}}$ near local noon to $\sim1500$--$2400~e^{-}\,\mathrm{s^{-1}\,pix^{-1}}$ at night. 
If uncorrected, however, such a dark current variation would introduce a systematic uncertainty of $0.08$--$0.12~\mathrm{mag~arcsec^{-2}}$ in the measured nighttime sky brightness.
To correct for this effect, we rescaled the laboratory dark frame using the warm-pixel scaling method \citep{ma2014new,ma2018first}, a method previously validated for AIRBT images by \citet{dong2025antarctic}, to match the dark-current distribution of each science frame.
The correction appears robust. Despite significant variations in the FPA temperature (e.g., from $-45.4^\circ$C to $-62.9^\circ$C on 16 April 2024; see Fig.~\ref{fig:sky_conti}), the derived sky brightness remains consistent, with no discontinuities or abrupt jumps that would indicate residual dark-current artifacts.
 
\subsection{Calibration and sky brightness measurement}

After image preprocessing, photometric calibration is performed using cross-matched stars within the field of view.
Source extraction and background estimation are carried out with \textsc{SExtractor} \citep{bertin1996sextractor}, and astrometric calibration is completed with \textsc{SCAMP} \citep{bertin2006automatic}.
The resulting internal source catalog is cross-matched with the 2MASS survey \citep{skrutskie2006two}.

For a matched star with 2MASS magnitude $m_\mathrm{2MASS}$ and measured flux \texttt{FLUX\_AUTO}, the zeropoint is computed as
\begin{equation}
m_\mathrm{zero} = m_\mathrm{2MASS} + 2.5 \log_{10} \left( \frac{\mathrm{counts}}{\mathrm{exp}} \right),
\end{equation}
where $counts$ is the stellar flux measured in ADU, and $exp$ is the exposure time in seconds.
The sky brightness is then obtained from
\begin{equation}
m_\mathrm{sky} = m_\mathrm{zero} - 2.5 \log_{10}
\left( \frac{\mathrm{counts}_{\mathrm{sky}} / \mathrm{exp}}{\mathrm{scale}^2} \right),
\end{equation}
where $counts_{\mathrm{sky}}$ is the measured sky level in ADU and $scale$ is the pixel scale in arcsec~pixel$^{-1}$.

It is important to note that the zeropoint derived from stellar photometry does not necessarily represent the appropriate calibration for sky-brightness measurements. 
The stellar zeropoint reflects not only the instrumental throughput but also additional attenuation caused by environmental conditions, such as cloud extinction and frosting on the entrance window. 
Under cloudy conditions, the stellar zeropoint can decrease significantly, but this does not imply that the sky background itself becomes brighter. 
Frost accumulation on the entrance window further complicates the calibration because it suppresses the throughput of point sources much more strongly than that of the diffuse sky background, whose photons can still reach the detector through scattering. 
Consequently, the real-time stellar zeropoint generally underestimates the intrinsic instrumental throughput. 

\begin{figure*}
\sidecaption
\includegraphics[width=12cm]{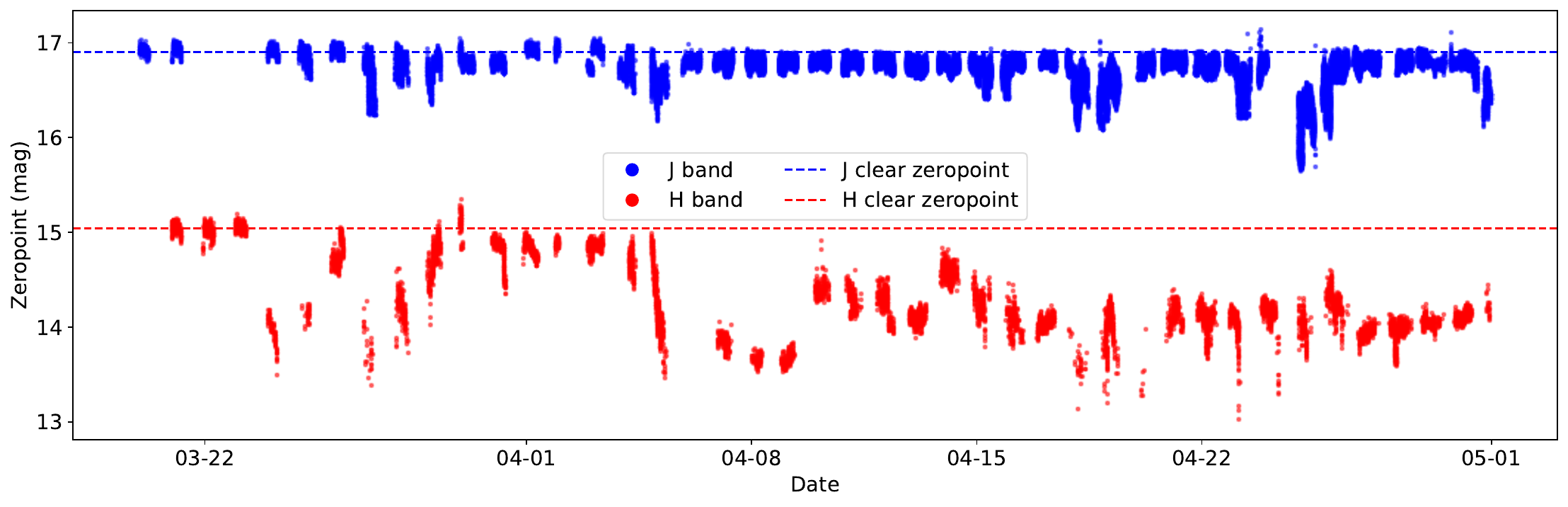}
\caption{
Temporal variations of the photometric zeropoint in the $J$ (blue) and $H$ (red) bands over the observing campaign.
The $J$-band zeropoint shows only a mild decline, whereas the $H$ band exhibits a clear and systematic throughput loss caused by frosting on the entrance window.
The adopted clear zeropoints for $J$ (blue dashed line) and $H$ (red dashed line) correspond to the median values measured during the first five clear nights and the first three clear nights, respectively.
}
\label{fig:zero}
\end{figure*}

To obtain an accurate sky brightness, we therefore adopt a clear zeropoint, defined under cloud-free and frost-free conditions. 
The clear zeropoint is determined from early observations when frosting is negligible. 
Because the two OTAs have different entrance-window heating power, the onset and severity of frosting differ between the $J$ and $H$ bands. 
As shown in Fig.~\ref{fig:zero}, the $J$-band zeropoint declines only mildly over the campaign, whereas the $H$ band, whose entrance-window heater provides only one-quarter of the $J$-band heating power, exhibits a clear and systematic throughput loss due to frosting. 
We therefore adopt the median zeropoint from the first five nights for $J$ and from the first three nights for $H$.

\begin{figure*}
\sidecaption
\includegraphics[width=12cm]{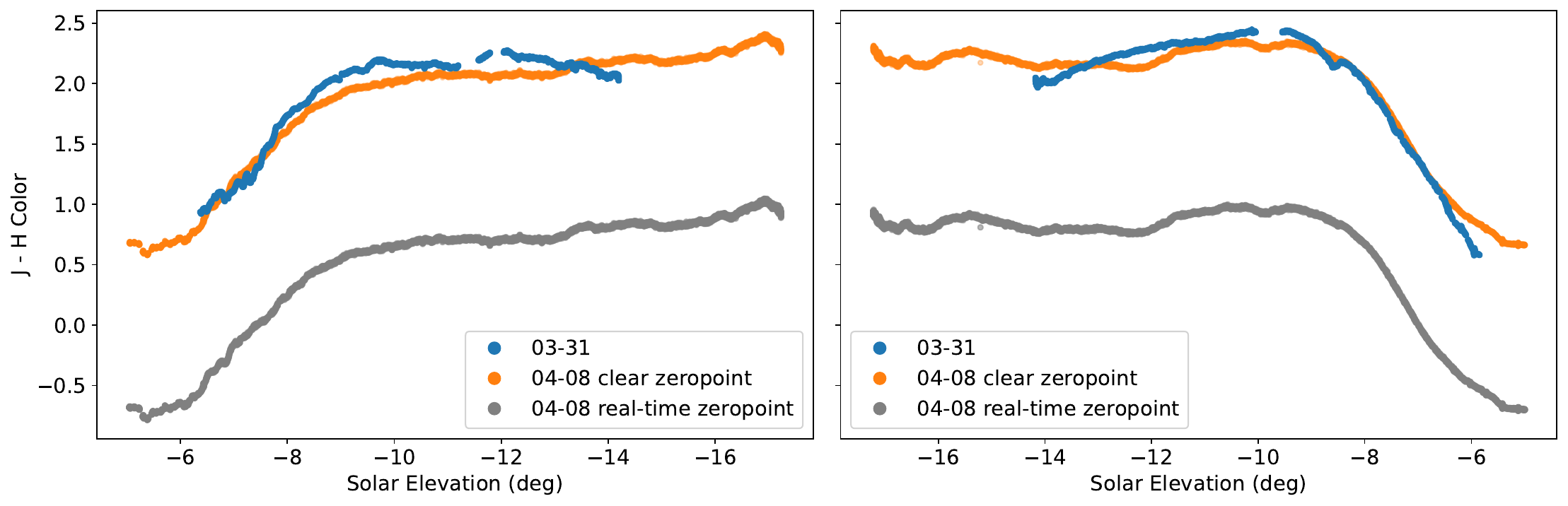}
\caption{
Demonstration of the robustness of the adopted clear zeropoint for deriving the sky background under frosting conditions.
Sky $J\!-\!H$ color on 31 March (blue, frost-free) and 8 April (affected by frosting) is shown as a function of solar elevation.
For 8 April, colors derived using the adopted clear zeropoint (orange) and those obtained with the real-time zeropoint (gray) are compared with the frost-free 31 March reference.
The colors derived with the clear zeropoint are much closer to the frost-free baseline, indicating that the clear zeropoint provides a more reliable estimate of the sky background than the real-time zeropoint.
}
\label{fig:JHcolor}
\end{figure*}

The efficacy of the clear zeropoint is demonstrated by comparing the $J\!-\!H$ sky color under frosting and frost-free conditions.
The $J$- and $H$-band sky backgrounds exhibit day-to-day variability, but because both bands are dominated by OH airglow, the resulting $J\!-\!H$ color remains a robust and repeatable quantity at a given solar elevation.
On 8 April, the $H$-band real-time zeropoint indicates that this was one of the most severely frosted nights of the campaign.
As shown in Fig.~\ref{fig:JHcolor}, the $J\!-\!H$ colors calculated using the adopted clear photometric zeropoint closely match those from the frost-free 31 March dataset, whereas the colors obtained using the real-time zeropoint differ substantially.

\subsection{Assessment of starlight}

In this section we assess the impact of integrated starlight on the sky background measurements.
The limiting magnitudes (2~s, $5\sigma$) of AIRBT are approximately 
11.5~mag in the $J$ band and 9.9~mag in the $H$ band. 
Using 2MASS catalogs of the observed fields, we further quantify the integrated starlight from sources fainter than the limiting magnitudes.
In a sparse field ($\sim$4500 stars), the contribution is 22.81~mag~arcsec$^{-2}$ in $J$ and 22.52~mag~arcsec$^{-2}$ in $H$.
In a crowded field ($\sim$17000 stars), the integrated brightness rises to 21.56~mag~arcsec$^{-2}$ in $J$ and 21.25~mag~arcsec$^{-2}$ in $H$. 
These values are still much fainter than the typical NIR sky brightness at Dome~A, indicating that the contribution of unresolved starlight to our sky background measurements is negligible.

\section{Results}
\label{sec:results}

\begin{figure*}
\centering
\includegraphics[width=17cm]{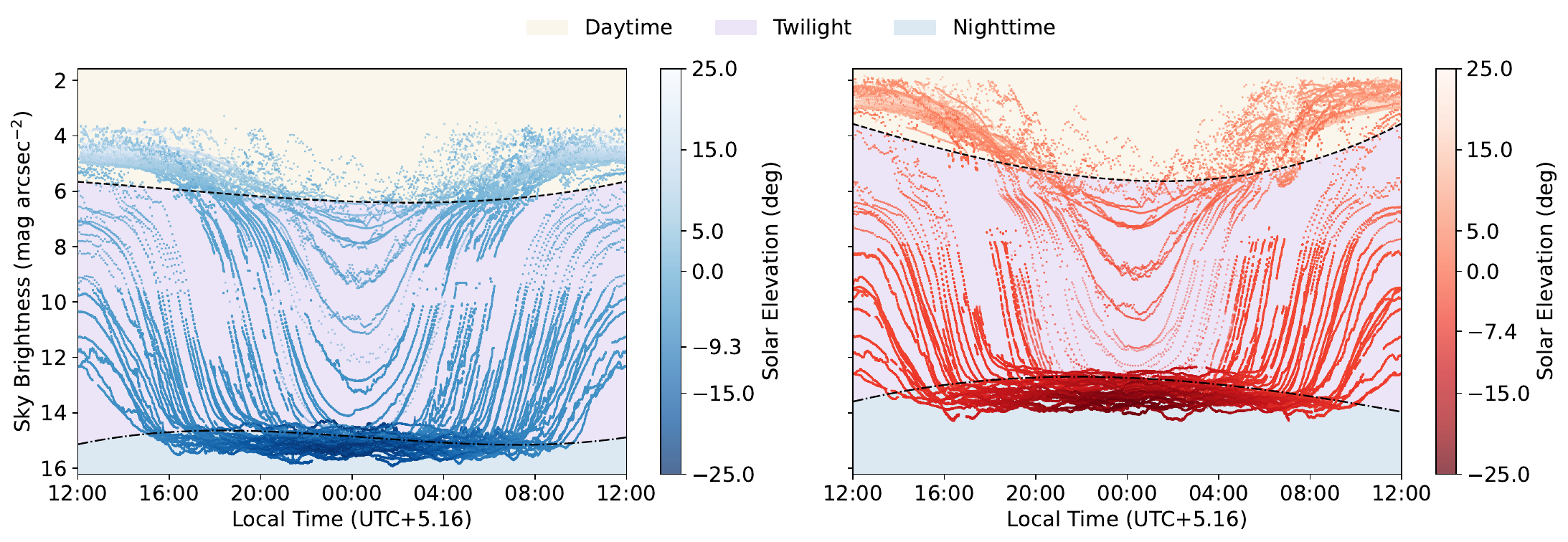}
\caption{
Overview of the NIR sky brightness measurements in the $J$ (left) and $H$ (right) bands during the observing campaign.
The shaded regions provide an approximate visual guide to the daytime (pale yellow), twilight (lavender), and nighttime (light blue) regimes.
The daytime--twilight boundary is set at a solar elevation of $0^\circ$ for both bands, while the twilight--nighttime boundary is set at $-9.3^\circ$ for $J$ and $-7.4^\circ$ for $H$, as discussed in Sect.~\ref{sec:Twilight}.
}
\label{fig:allday}
\end{figure*}

Our $J$- and $H$-band monitoring at Dome~A characterizes the NIR sky background from the end of the polar day through the onset of the polar night. 
Figure~\ref{fig:allday} presents an overview of the sky brightness measurements obtained during the 2024 observing campaign.
The wide range of sampled solar elevations allows us to examine both diurnal and seasonal variations in the NIR sky background.

Based on the variation of the $J$- and $H$-band sky brightness with solar elevation, the observing period can be divided into astronomical daytime, twilight and nighttime for each band.

\subsection{Daytime}

Daytime is defined as periods with solar elevation above $0^\circ$, which persist until 15 April during the 2024 observing season at Dome~A.
The median daytime sky brightness is 5.2 and 2.9~mag~arcsec$^{-2}$ in the $J$ and $H$ bands, respectively. 
Over the daytime solar elevation range of $0^\circ$--$25^\circ$, the sky brightness in both bands varies only gradually with solar elevation.
\citet{yang2025first} reported $J$-band sky brightness values of 2.95--5.78~mag~arcsec$^{-2}$ from selected fields observed in January--February 2024, when the solar elevation was generally higher than during our observing period. 
Their measurements therefore extend toward brighter daytime conditions but remain broadly consistent with the daytime brightness range determined in this work.

\subsection{Twilight}
\label{sec:Twilight}
\begin{figure*}
\sidecaption
\includegraphics[width=12cm]{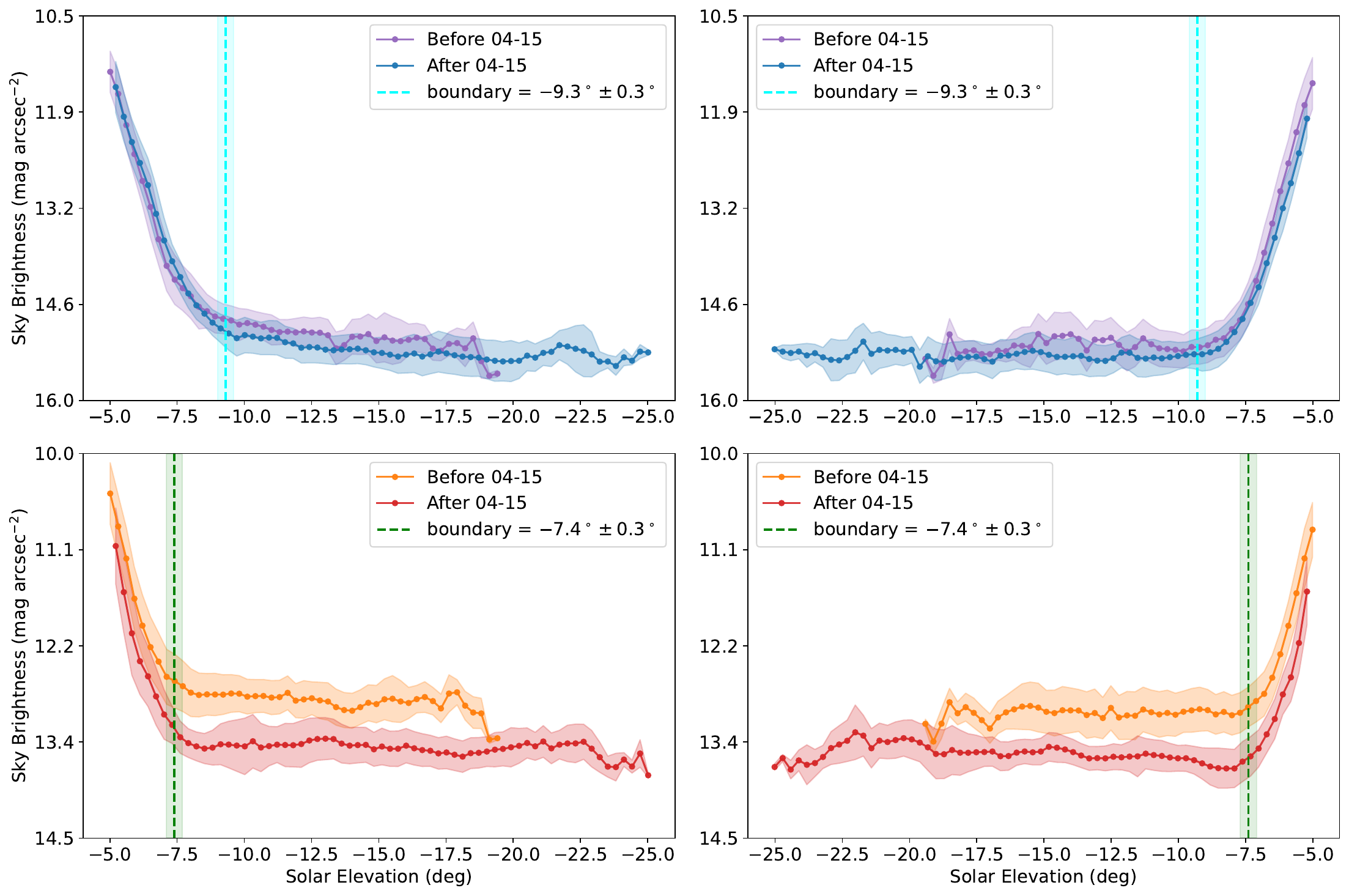}
\caption{
Sky brightness as a function of solar elevation in the $J$ (upper) and $H$ (lower) bands for dusk (left) and dawn (right).
Points show the median sky brightness in sliding bins of solar elevation, and the shaded regions indicate the standard deviation.
Vertical shaded bands mark the twilight--nighttime transition derived from the dawn branch.
Data obtained before and after 15 April are shown separately, as this date marks the beginning of the polar night at Dome~A.
}
\label{fig:twilight}
\end{figure*}

During twilight, the $J$- and $H$-band sky brightness varies rapidly.
Fig.~\ref{fig:twilight} shows the evolution of the sky background during the dusk and dawn transitions.
After dusk, however, the sky continues to fade gradually for an extended period.
To avoid this asymmetry, we therefore use only the dawn branch to define the boundary between the relatively stable nighttime regime and the rapidly varying twilight regime.

We adopt the median sky brightness around local midnight as the reference level for the nighttime background.
Along the dawn branch, the sky brightness becomes systematically higher than the value corresponding to this reference level at a given solar elevation, marking the transition between the nighttime and twilight regimes.
The twilight--nighttime transition occurs at solar elevations of $-9.3^\circ \pm 0.3^\circ$ in the $J$ band and $-7.4^\circ \pm 0.3^\circ$ in the $H$ band.
Within these uncertainties, the boundary angles derived before and during the polar night (which begins on 15 April) are consistent.

For comparison, the results of \citet{zhang2023sky} indicate twilight--nighttime transition elevations of $-7.8^\circ$ in $J$ and $-6.2^\circ$ in $H$, both higher than the values derived in this work.
This difference may be attributed to three factors. 
First, the transition is defined differently. \citet{zhang2023sky} used a model-dependent knee point, whereas our definition is data-driven and based on the approach to the nighttime baseline.
Second, their analysis relied on a more limited dataset and did not separate the dusk and dawn branches. 
Third, their observations were conducted during the solar minimum, when OH airglow is expected to be weaker, causing the sky to become nighttime-dominated at a higher solar elevation.

\subsection{Nighttime}
\label{sec:nighttime}
During the astronomical nighttime, a clear separation is observed between the periods before 15 April and the polar-night period (see Fig.~\ref{fig:twilight}).
During the polar-night period, the nighttime sky background is lower by about $0.1$ and $0.4$~mag~arcsec$^{-2}$ in the $J$ and $H$ bands, respectively, when compared at the same solar elevations.

During the dusk branch, both the $J$- and $H$-band sky brightness exhibit an initial decrease at the beginning of nighttime, becoming dimmer by $\sim0.4$~mag~arcsec$^{-2}$ over $\sim1.5$~hours in $J$ and $\sim0.3$~mag~arcsec$^{-2}$ over $\sim0.5$~hours in $H$.
This effect is much stronger at mid-latitude sites such as Mauna Kea and Lenghu, Qinghai \citep{roth2016measurements, li2024preliminary}, where the $J$-band sky background continues to decline by nearly $\sim1$~mag~arcsec$^{-2}$ after the solar elevation has dropped below $-12^\circ$, over a duration of 1.78--3.36~hours.
In contrast, at Dome~A the decrease in both $J$- and $H$-band sky brightness is significantly smaller and much shorter in duration.
This indicates that once the site has fully entered nighttime, the NIR sky background at Dome~A remains significantly more stable than at typical mid-latitude observatories.
Such stability reduces long-term background variations, offering clear advantages for deep NIR imaging and for high-precision time-domain observations of faint sources.

Based on the derived transition boundaries, we estimate the annual nighttime hours in the $J$ and $H$ bands. 
As summarized in Tab.~\ref{tab:sun_hours_sites}, Dome~A provides about $3219$~hours and $3493$~hours of nighttime in the $J$ and $H$ bands, respectively, compared to about $2600$~hours in the optical $i$ band. 
This corresponds to an additional $620$--$890$~hours ($25$--$35\%$) in the NIR. 
Because Dome~A reaches the nighttime sky-background level at significantly higher solar elevations than the conventional $-12^\circ$ criterion used at mid-latitude sites, the resulting nighttime windows in the $J$ and $H$ bands are comparable to those at mid-latitude observatories despite the extreme latitude of the site. 
Considering the substantially higher clear-sky fraction at Dome~A \citep{yang2021cloud}, the effective NIR observing time is expected to exceed that at mid-latitude sites, whereas the corresponding optical observing time is noticeably shorter.

\begin{table*}[htbp]
\centering
\caption{
Annual nighttime hours in 2024 at several major astronomical sites.
Nighttime boundaries are defined by solar elevations of $-13^\circ$ (optical, Dome~A), $-9.3^\circ$ ($J$, Dome~A), $-7.4^\circ$ ($H$, Dome~A), $-18^\circ$ (optical, mid-latitude sites), and $-12^\circ$ (near-infrared, mid-latitude sites).
}
\begin{tabular}{lcccc}
\hline
Site & Latitude & Elevation (m) & Optical & Near-infrared \\
\hline
Dome A (Antarctica)      & $-80^\circ22'$ & 4093 & 2603 & 3219 ($J$), 3493 ($H$) \\
Mauna Kea (Hawaii, USA)  & $+19^\circ50'$ & 4194 & 3670 & 3908 \\
Cerro Pachón (Chile)     & $-30^\circ14'$ & 2722 & 3631 & 3805 \\
La Palma (Spain)         & $+28^\circ46'$ & 2332 & 3601 & 3852 \\
Lenghu (China)           & $+38^\circ36'$ & 4000 & 3487 & 3729 \\
\hline
\end{tabular}
\label{tab:sun_hours_sites}
\end{table*}

\begin{figure}
\resizebox{\hsize}{!}{\includegraphics{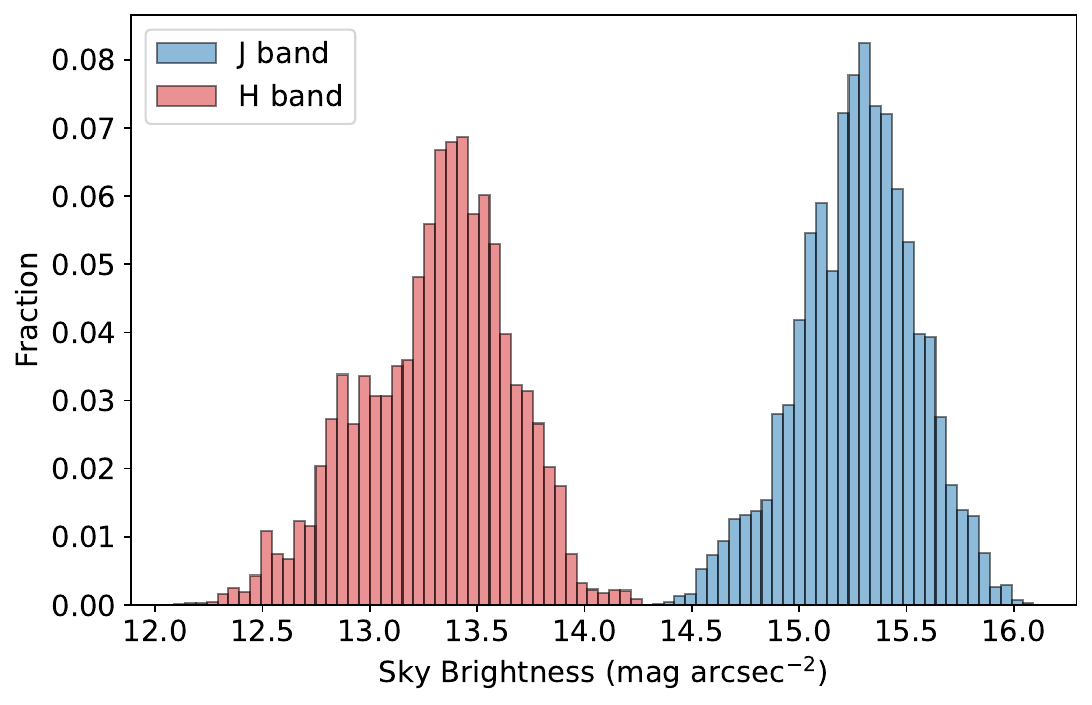}}
\caption{Nighttime sky brightness histograms for the $J$ (blue) and $H$ (red) bands.}
\label{fig:hist}
\end{figure}

A total of 41 nights of usable data were obtained in both the $J$ and $H$ bands, yielding median sky brightnesses of $15.3~\mathrm{mag~arcsec^{-2}}$ in $J$ and $13.4~\mathrm{mag~arcsec^{-2}}$ in $H$.
Fig.~\ref{fig:hist} presents the histogram distributions of the nighttime sky background for both bands.
These values are compared with those reported for other major astronomical sites worldwide in Tab.~\ref{tab:site_comparison}. 
Our measurements are noticeably brighter than previously reported NIR sky-brightness measurements for Antarctic sites, a point we discuss further in Sect.~\ref{sec:discussion}.

\begin{table*}[htbp]
\centering
\caption{Nighttime NIR sky brightness reported at several major astronomical sites.}

\begin{minipage}{\linewidth}
\centering
\begin{tabular}{lcccc}
\hline
Site & Type & $J$ band (mag~arcsec$^{-2}$) & $H$ band (mag~arcsec$^{-2}$) & References \\
\hline
Dome A (Antarctica)$^{a}$ & Median & 15.3 & 13.4 & This work \\
Dome A (Antarctica)$^{b}$ & -- & 16.0--15.35 & 14.85--13.92 & \citet{zhang2023sky} \\
South Pole (Antarctica) & -- & 16.8--16.0 & 15.2--14.2 & \citet{phillips1999near} \\
Mauna Kea (Hawaii, USA) & Average & 15.6 & 14.0 & \citet{sanchez2008night} \\
La Palma (Spain) & Average & 15.5 & 14.0 & \citet{sanchez2008night} \\
Cerro Pachón (Chile) & -- & 16.0 & 13.9 & \citet{sanchez2008night} \\
Lenghu (China)$^{c}$ & Median & 15.6 & 13.8 & \citet{li2024preliminary} \\
Mount Abu (India) & Average & 15.23 & 13.76 & \citet{prajapati2023near} \\
Caucasus Mountain (Russia) & Average & 15.7 & 13.9 & \citet{tatarnikov2024brightness} \\
\hline
\end{tabular}

\vspace{2mm}

\raggedright
\textbf{Notes.}
$^{a}$ Early polar-night measurements obtained during solar maximum; $H$ is $\sim0.11$~mag~arcsec$^{-2}$ brighter than 2MASS due to bandpass mismatch (no correction applied). \\
$^{b}$ Early polar-night measurements obtained during solar minimum. \\
$^{c}$ Preliminary results based on only a few months of observations; $H$ is $\sim0.11$~mag~arcsec$^{-2}$ brighter than 2MASS due to bandpass mismatch (no correction applied).
\end{minipage}

\label{tab:site_comparison}
\end{table*}

\subsection{Zenith cloud coverage}

\begin{figure*}
\centering
\includegraphics[width=17cm]{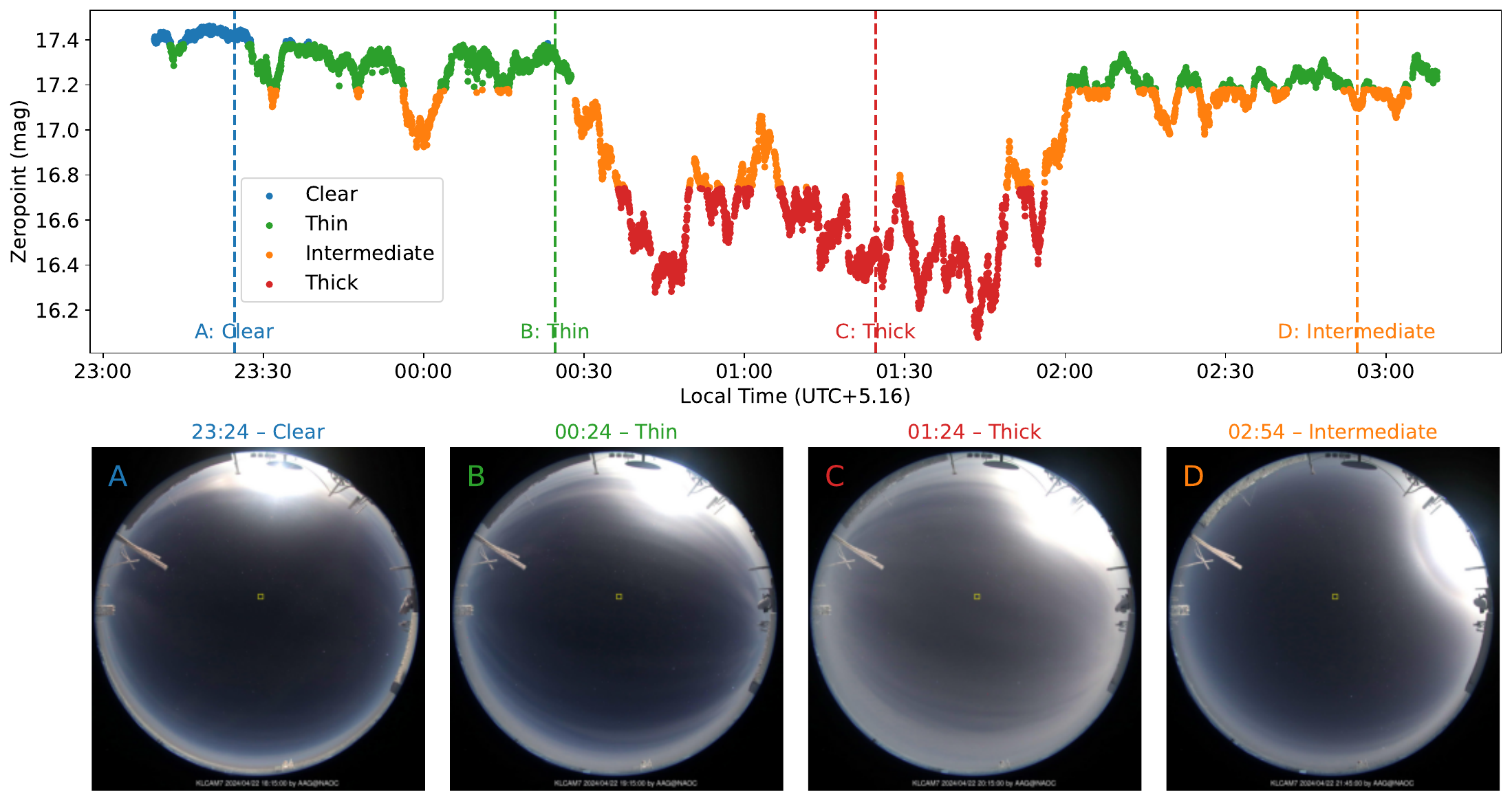}
\caption{
Comparison between cloud conditions inferred from the AIRBT $J$-band photometric zeropoint and the KLCAM all-sky optical images.
The upper panel shows the temporal variations of the zeropoint on 22 April, with atmospheric transparency classified into the categories of Clear (blue), Thin (green), Intermediate (orange), and Thick (red), respectively.
The lower panel displays the corresponding KLCAM all-sky images at the times indicated by the vertical lines.
The yellow boxes mark the AIRBT observing field.
Overall, the cloud patterns inferred from the zeropoint variations are broadly consistent with those visible in the KLCAM all-sky images.
}
\label{fig:zp}
\end{figure*}

In addition to the sky background measurements, we also use variations in the $J$-band photometric zeropoint to evaluate the cloud conditions near zenith.
To ensure the reliability of the zeropoint-based cloud estimates, we first compare them with the all-sky optical images obtained by KLCAM \citep{yang2021cloud}. As shown in Fig.~\ref{fig:zp}, the cloud patterns inferred from the zeropoint variations are generally consistent with those visible in the KLCAM all-sky images, demonstrating that the $J$-band zeropoint provides a reliable indicator of zenith transparency.

As shown in Tab.~\ref{tab:cloud}, approximately 72.8\% of the measurements exhibit relative transparency above 90\%, compared with about 51\% reported from CSTAR $i$-band data \citep{zou2010sky}.
Part of this difference likely arises because infrared wavelengths are less sensitive to extinction by thin clouds. 
In addition, the much smaller field of view of AIRBT ($\sim1~\mathrm{deg}^2$) compared with CSTAR ($\sim20~\mathrm{deg}^2$) makes it less likely to sample patchy cloud structures, which can lead to systematically higher transparency estimates.
Consequently, transparency values inferred from the $J$-band zeropoint should be regarded as a proxy for cloud fraction, although they may be systematically biased toward higher transparency.

\begin{table}[htbp]
\centering
\caption{Cloud-cover statistics derived from AIRBT, compared with CSTAR results \citep{zou2010sky}.}
\begin{tabular}{ccc c}
\hline
\begin{tabular}{c} Flux \\ \& Extinction (mag) \end{tabular} 
& AIRBT & CSTAR & Cloud Cover \\
\hline
$> 90\%$ \ ($<0.11$)      & 72.8\% & 51\% & Clear \\
75\%--90\% \ (0.11--0.31) & 15.3\% & 23\% & Thin \\
50\%--75\% \ (0.31--0.75) & 10.4\% & 17\% & Intermediate \\
$< 50\%$ \ ($>0.75$)      & 1.5\%  & 9\%  & Thick \\
\hline
\end{tabular}
\label{tab:cloud}
\end{table}

\section{Discussion}
\label{sec:discussion}

The nighttime NIR sky background at Dome~A in 2024 is measurably brighter than that reported in previous Antarctic campaigns and at other major astronomical sites.
In this section, we investigate the physical origins of this difference by examining the potential influence of solar activity, lunar illumination, and auroral emission on the observed $J$- and $H$-band sky background.

\subsection{Solar and geomagnetic activities}

\begin{figure}
  \resizebox{\hsize}{!}{\includegraphics{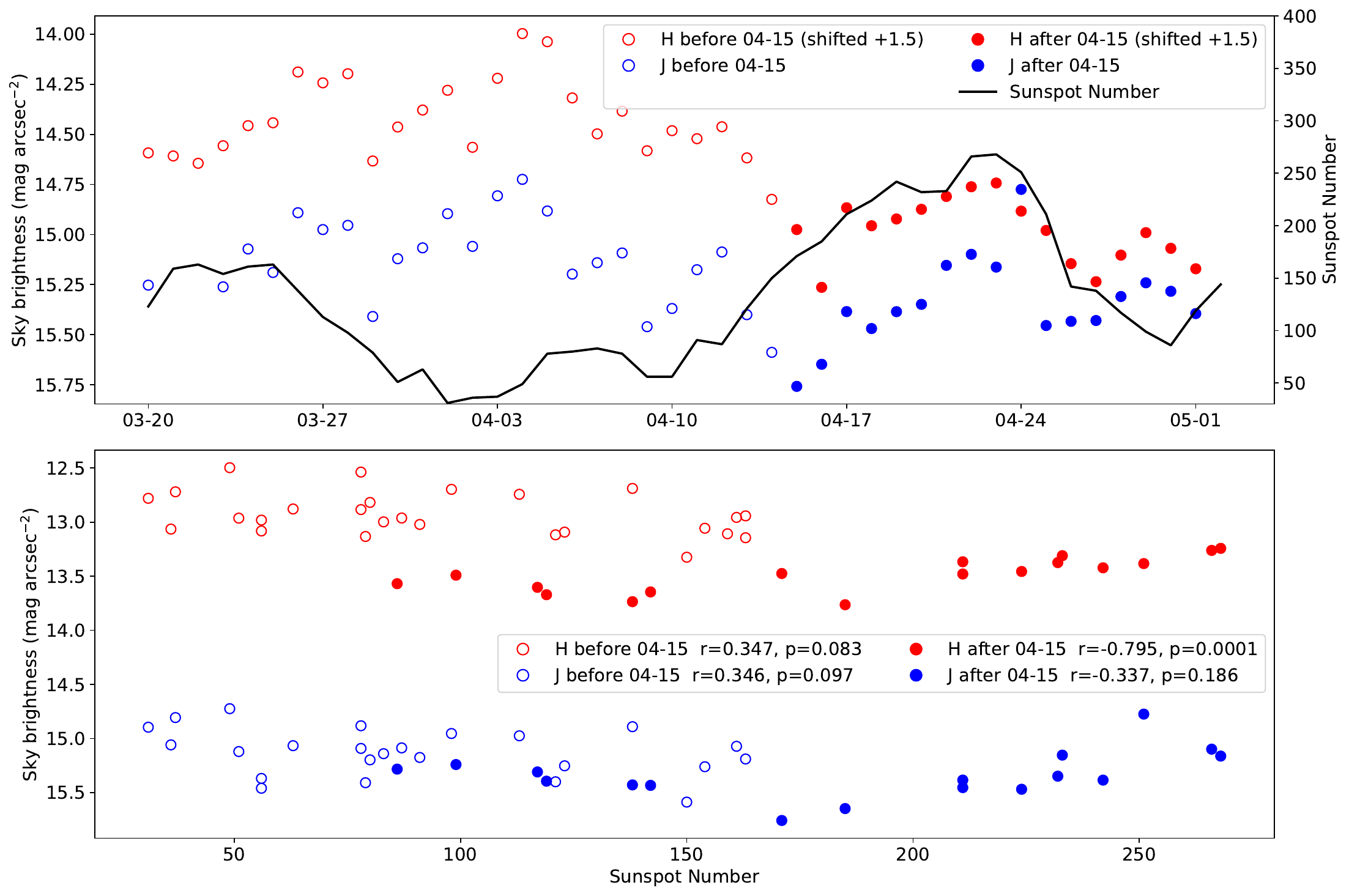}}
  \caption{
  Top: Daily median sky brightness in the $J$ (blue) and $H$ (red, shifted downward by 1.5~mag) bands compared with the variation of sunspot number (black line).
  Bottom: Relationship between nighttime sky brightness and sunspot number. 
  Open circles represent measurements obtained before the onset of the polar night, while filled circles indicate data taken after the onset of the polar night (post--15 April).
  During 16--23 April, the $J$- and $H$-band sky backgrounds show temporal variations broadly similar to those of the sunspot number. 
  Using all data obtained after the onset of the polar night, the Spearman rank analysis gives $r_s=-0.795$ ($p=0.0001$) for the $H$ band and $r_s=-0.337$ ($p=0.186$) for the $J$ band.
  }
  \label{fig:spot}
\end{figure}

Based on observations obtained between 20 March and 1 May 2024, we investigate the 
relationship between the nighttime sky brightness and sunspot number 
(\citealt{SILSO_SN_V2}) before and after the onset of the polar night 
(Fig.~\ref{fig:spot}). 
The sky brightness is characterized using the daily median value.
On timescales of days, the sky background before the onset of the polar night shows 
no significant correlation with sunspot activity. 
After the onset of the polar night, however, the $J$- and $H$-band sky backgrounds during 16--23 April show temporal variations broadly similar to those of the sunspot 
number, especially in the $H$ band.
Our preliminary interpretation is that, during the polar night period, the nighttime sky background variability may be partly related to sunspot activity.
As a quantitative check, we performed a Spearman rank analysis for the polar-night data, obtaining $r_s = -0.795$ ($p = 0.0001$) for the $H$ band and 
$r_s = -0.337$ ($p = 0.186$) for the $J$ band.
However, two caveats should be noted. 
First, within the polar night data, the NIR sky background appears to increase with sunspot number only when the sunspot number exceeds a certain threshold. 
Second, the daily median sky brightness also shows additional short-timescale variability, including local enhancements on a timescale of approximately one week. 
Such variability may contribute to cases where the sky brightness and sunspot number 
do not evolve synchronously.
The current data set is not sufficiently long to robustly determine this threshold, 
and longer-term monitoring will be required to confirm and characterize this behavior.

On annual timescales, the optical night sky is known to be modulated by the 11-year solar cycle, which can be traced by the 10.7~cm radio flux or sunspot number. 
Although optical sky brightness does not generally respond to sunspot variations on short timescales, the mean $V$-band sky level between solar maximum and minimum can differ by up to $\sim$0.5~mag \citep{walker1988effect,krisciunas1997optical,alarcon2021natural}.
Based on the above discussion, it is plausible that the NIR background may be 
brighter at higher sunspot numbers during the polar-night period.
This expectation is supported by the fact that near-infrared night-sky emission is dominated by airglow driven by solar ultraviolet radiation, which responds to variations in solar activity \citep{noll2025palace}.
Previous Antarctic NIR measurements, including those obtained at the South Pole in 1995 and at Dome A in 2019, were collected near solar minimum, whereas our 2024 observations were obtained close to solar maximum.
Therefore, the higher sky brightness reported here is not surprising and provides a reference baseline for future comparative measurements at different phases of the solar cycle.

\begin{figure}
\resizebox{\hsize}{!}{\includegraphics{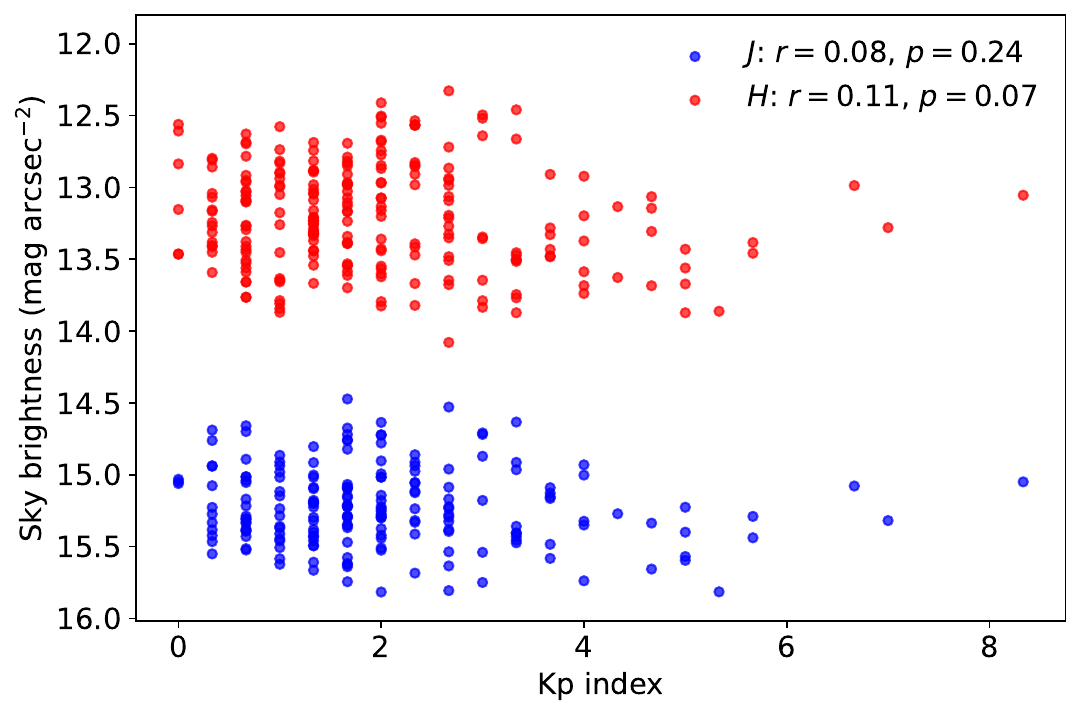}}
\caption{
Nighttime sky brightness in the $J$ (blue) and $H$ (red) bands as a function of the 3-hourly Kp index.
}
\label{fig:kp}
\end{figure}

Previous studies have suggested a potential link between optical sky brightness and geomagnetic activity \citep{grauer2019impact}.
We therefore examined the relationship between NIR sky brightness and geomagnetic activity using the 3-hourly Kp index.
For each Kp interval, we calculated the median nighttime sky brightness from the corresponding 3-hour observing period.
As shown in Fig.~\ref{fig:kp}, no significant dependence on the Kp index is found in either band.
The Spearman rank coefficients are $r_s=0.08$ ($p=0.24$) for the $J$ band and $r_s=0.11$ ($p=0.07$) for the $H$ band.
These results suggest that geomagnetic disturbances are unlikely to be the dominant driver of the observed short-timescale NIR sky-background variations.

\subsection{Effect of the Moon}

Previous studies at mid-latitude observatories have shown that the NIR sky background in the $J$, $H$, and $K$ bands is essentially insensitive to moonlight, regardless of the lunar phase or the angular distance between the Moon and the observed field \citep{roth2016measurements}. 
Under the unique geometric conditions of Dome~A, the Moon never rises very high above the horizon. 
Its maximum elevation reaches only $\sim39^\circ$, corresponding to a minimum angular distance of $\sim51^\circ$ from the zenith.
Given this large separation between the Moon and the zenith, scattered moonlight is therefore expected to have only a negligible influence on the zenith sky background in both the $J$ and $H$ bands.

\begin{figure}
\resizebox{\hsize}{!}{\includegraphics{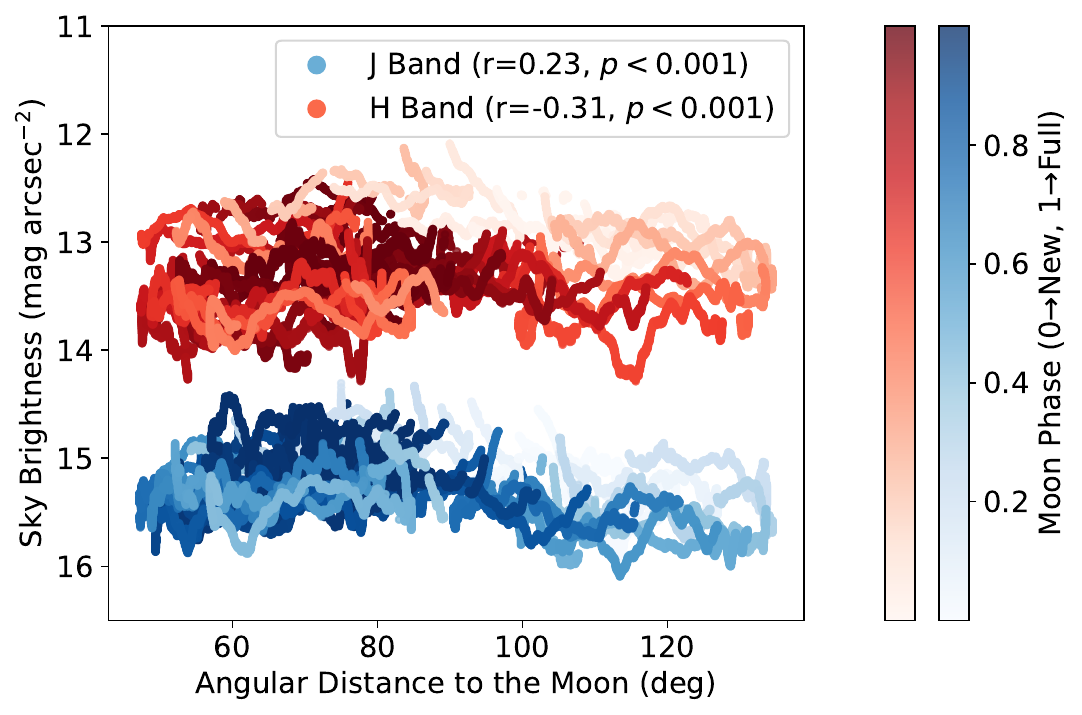}}
\caption{
Correlation between the $J$- and $H$-band sky brightness and the angular distance to the Moon.
Individual measurements are shown, with the color scale indicating lunar phase from 0 (new Moon) to 1 (full Moon).
}
\label{fig:moon}
\end{figure}

Motivated by this, we quantitatively evaluated the potential lunar contribution at Dome~A by performing a correlation analysis between nighttime sky brightness and the angular distance to the Moon. 
As shown in Fig.~\ref{fig:moon}, the Spearman rank correlation coefficients are $r_s = 0.23$ ($p \ll 0.001$) in the $J$ band and $r_s = -0.31$ ($p \ll 0.001$) in the $H$ band.
Owing to the large number of data points ($>5\times10^{5}$), the very small $p$ values reflect high statistical power rather than a strong correlation.
In Fig.~\ref{fig:moon}, the sky-background measurements appear to show some separation between different lunar phases. 
However, this apparent phase dependence is likely affected by sampling bias between the pre-polar-night and polar-night observations. 
Low-phase measurements at large lunar angular distances are mainly obtained before the onset of the polar night, whereas high-phase measurements are dominated by polar-night observations. 
Since the nighttime sky background differs between these two periods 
(Sect.~\ref{sec:nighttime}), the uneven sampling can mimic a lunar-phase dependence. 
We therefore do not find clear evidence that lunar phase or angular distance to the
Moon is the dominant driver of the observed NIR sky-background variability.

\subsection{Aurora}

Most studies of auroral emission have focused on the optical regime, where the strongest emission lines occur at wavelengths shorter than $\sim0.9~\mu$m \citep{sims2012airglow,petrie1950near}. 
Whether auroral activity has a measurable impact on the NIR sky background in the $J$ and $H$ bands remains poorly constrained by observations.

\begin{figure*}
\centering
\includegraphics[width=17cm]{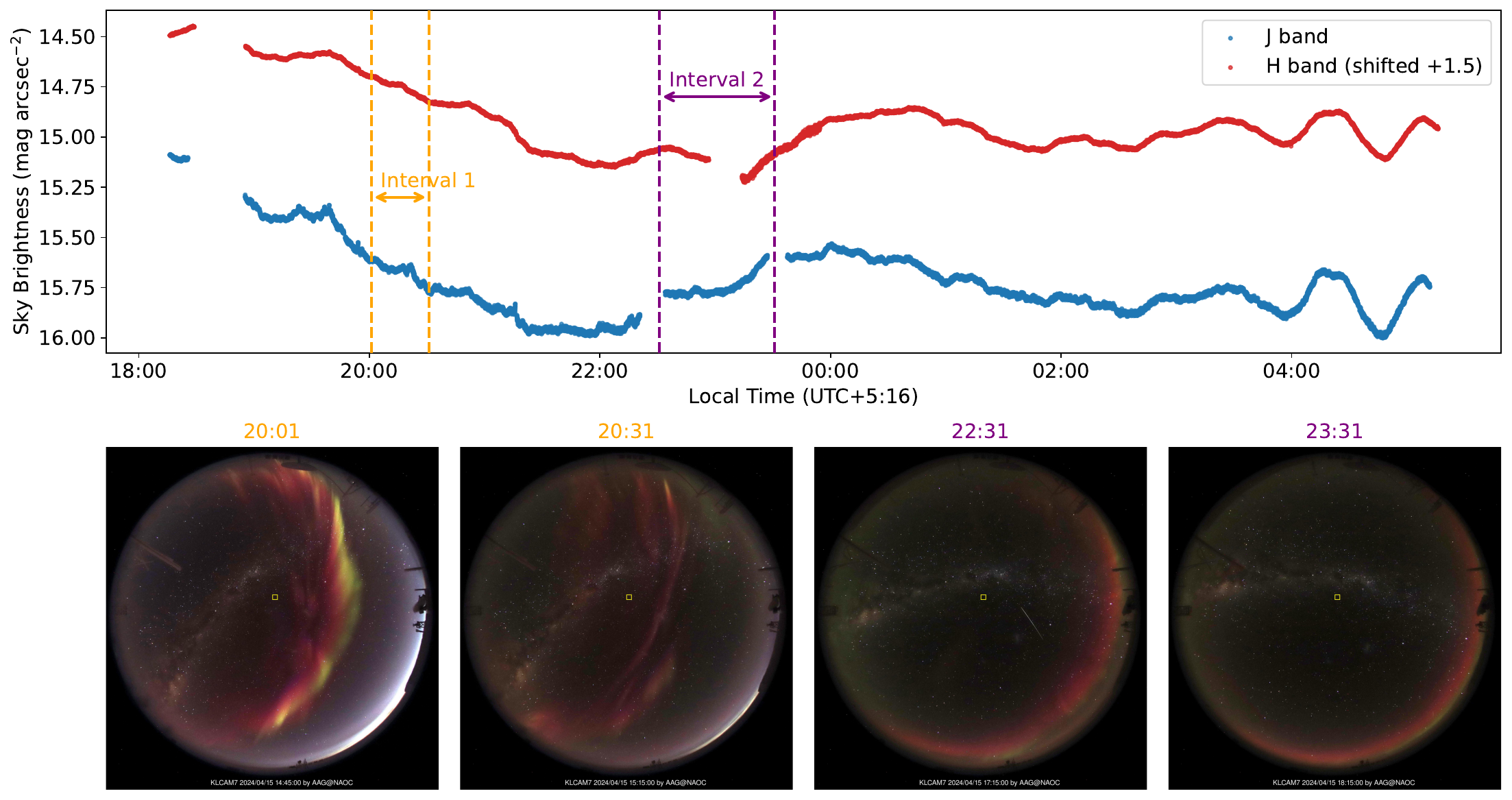}
\caption{
Comparison between the $J$- and $H$-band sky background and the optical all-sky auroral images.
The upper panel shows the $J$-band sky brightness (blue) and the $H$-band sky brightness (red, shifted downward by 1.5~mag) on 15 April.
The lower panel displays the corresponding KLCAM all-sky images, with the AIRBT observing field indicated by yellow boxes.
During Interval~1, strong auroral emission is clearly visible in the optical all-sky images, while the NIR sky background in both the $J$ and $H$ bands continues to decrease.
In contrast, during Interval~2 the $J$- and $H$-band sky brightness increases, whereas the optical auroral activity appears comparatively weak.
These comparisons indicate that the variations in the NIR sky background are not directly associated with the optical auroral activity.
}
\label{fig:aurora}
\end{figure*}

To explore this possibility, we examined periods during which KLCAM all-sky images show strong auroral contamination in the optical bands, and compared them with the contemporaneous $J$- and $H$-band sky brightness variations.
As illustrated in Fig.~\ref{fig:aurora}, during Interval~1 strong auroral emission is clearly visible in the optical all-sky images, while the $J$- and $H$-band sky background continues to decrease. 
In contrast, during Interval~2 the $J$- and $H$-band sky brightness increases, whereas the optical auroral activity appears comparatively weak. 
These results therefore suggest that optical auroral events are unlikely to be a major contaminant of the $J/H$-band sky background. 
A more definitive assessment, however, will require dedicated NIR spectroscopic observations of the aurora at Dome~A.

\subsection{Implications for future NIR observations at Dome A}

The measured $J$- and $H$-band sky backgrounds provide practical constraints for future NIR instrument design and science planning at Dome~A. 
They can be used to estimate the sky-background electron rate, set requirements on detector dark current, and evaluate the observing depth. 
As an illustrative estimate, we scale from the current AIRBT system, assuming the same pixel size and total throughput for future instruments.
The $H$-band values refer to the narrower InGaAs effective $H$ band (1.48--1.65~$\mu$m). 
The limiting magnitudes are estimated using the median nighttime sky background measured in this work; adopting the South Pole dark-end values in Table~\ref{tab:site_comparison} as a darker Antarctic reference would improve the limits by about 0.5~mag.

A 40-cm $f/2.5$ wide-field NIR telescope has been developed and is planned for installation at Dome~A in 2028.
For this system, the corresponding nighttime sky-background rates are $\sim2220$ and $\sim2720~e^-\,\mathrm{s^{-1}\,pix^{-1}}$ in the $J$ and $H$ bands, respectively.
Requiring the detector dark current to remain below 25\% of the sky background gives $I_{\rm dark}\lesssim560$ and 
$\lesssim680~e^-\,\mathrm{s^{-1}\,pix^{-1}}$ in $J$ and $H$, respectively, which is achievable with off-the-shelf InGaAs cameras. 
Assuming 20-s exposures and sky-background-dominated noise, the estimated $5\sigma$ limiting magnitudes are $J/H\simeq15.6/13.6$~mag.
This depth is suitable for high-cadence NIR time-domain surveys over large fields, including searches for variable sources and long-term monitoring of large samples.

A 1-m $f/8$ NIR telescope is also being planned to exploit the excellent free-atmosphere seeing at Dome~A.
For this slower system, the lower sky-background rates 
($\sim217/266~e^-\,\mathrm{s^{-1}\,pix^{-1}}$ in $J/H$) tighten the dark-current requirement to $I_{\rm dark}\lesssim54/67~e^-\,\mathrm{s^{-1}\,pix^{-1}}$.
With 180-s exposures, the estimated $5\sigma$ limits reach $J/H\simeq20.0/18.1$~mag.
This depth would be well suited to the follow-up of transients, such as supernovae, tidal disruption events, and high-redshift gamma-ray burst afterglows, as well as to high-precision photometry of variable targets, including variable stars, brown dwarfs, ultracool stars, and reddened sources.

\section{Summary}
\label{sec:summary}

In 2024, AIRBT performed the first continuous zenith near-infrared sky background monitoring at Dome~A, covering the transition from the end of the polar day to the onset of the polar night. A total of 73 days of $J$-band data and 75 days of $H$-band data were obtained. The median sky brightness is $5.2~\mathrm{mag~arcsec^{-2}}$ in $J$ and $2.9~\mathrm{mag~arcsec^{-2}}$ in $H$ during astronomical daytime, and $15.3~\mathrm{mag~arcsec^{-2}}$ in $J$ and $13.4~\mathrm{mag~arcsec^{-2}}$ in $H$ during nighttime. The twilight--nighttime transition occurs at solar elevations of $-9.3^\circ$ in $J$ and $-7.4^\circ$ in $H$. Consequently, Dome~A provides approximately $25\%$ and $35\%$ more observable nighttime in the $J$ and $H$ bands than in the optical.

A clear difference is observed between the pre--polar-night and polar-night periods. 
At the same solar elevation, the NIR sky background during the polar night is systematically darker by about $0.1$ and $0.4~\mathrm{mag~arcsec^{-2}}$ in the $J$ and $H$ bands, respectively, compared to the pre--polar-night period.
Compared with mid-latitude observatories, the decrease in sky brightness after sunset at Dome~A is both smaller in amplitude and shorter in duration, indicating a more stable NIR sky background once nighttime is established.

During the polar-night period, the nighttime sky background shows a tentative association with solar activity, more clearly in the $H$ band than in the $J$ band. 
This suggests that the NIR sky background may be partly affected by solar activity, 
although longer-term monitoring is required to confirm this possibility. 
Moreover, no meaningful correlation is found between the NIR sky background and either lunar distance or optical auroral activity.

Further progress will require long-term monitoring, particularly datasets that cover the full polar-night period and span different phases of the solar activity cycle, in order to establish a more complete picture of the temporal behavior of the Antarctic NIR sky background.

\begin{acknowledgement}
We acknowledge the support of the 39th, 40th, and 41st Chinese National Antarctic Research Expedition (CHINARE) teams, organized by the Polar Research Institute of China and the Chinese Arctic and Antarctic Administration. 
The Antarctic Infrared Binocular Telescope (AIRBT) project is supported by the School of Physics and Astronomy at Sun Yat-sen University. 
This work is also supported by the National Astronomical Observatories, Chinese Academy of Sciences under grant numbers E355350101 and E4TG350101, and by the National Natural Science Foundation of China under grant numbers 11733007 and 12373092.
\end{acknowledgement}

\bibliographystyle{aa}
\bibliography{ref.bib}
\end{document}